\documentclass[aps,prl,amsmath,amssymb,floatfix,twocolumn,amsmath,superscriptaddress,twocolumn,nofootinbib,tighten,letterpaper]{revtex4}
\usepackage{multirow}
\usepackage{bbold}
\usepackage{subfigure}
\usepackage{color}
\usepackage{mathrsfs}
\usepackage{hyperref}
\usepackage[normalem]{ulem}
\usepackage{bm}
\usepackage{pifont}

\renewcommand\vec[1]{\ensuremath\boldsymbol{#1}} 

\usepackage{amsfonts, relsize, color}
\usepackage{graphics}
\usepackage{graphicx}
\usepackage{subfigure}
\usepackage{hyperref}
\usepackage{color}

\begin{document}
\title{Mixed-parity octupolar pairing and corner Majorana modes in three dimensions}

\author{Bitan Roy}\thanks{Corresponding author: bitan.roy@lehigh.edu}
\affiliation{Department of Physics, Lehigh University, Bethlehem, Pennsylvania, 18015, USA}

\author{Vladimir Juri\v ci\' c}
\affiliation{Nordita, KTH Royal Institute of Technology and Stockholm University, Roslagstullsbacken 23,  10691 Stockholm,  Sweden}
\affiliation{Departamento de F\'isica, Universidad T\'ecnica Federico Santa Mar\'ia, Casilla 110, Valpara\'iso, Chile}

\date{\today}
\begin{abstract}
We identify time-reversal symmetry breaking mixed-parity superconducting states that feature eight Majorana corner modes in properly cleaved three-dimensional cubic crystals. Namely, when an odd-parity isotropic $p$-wave pairing coexists with cubic symmetry preserving even-parity octupolar $d_{x^2-y^2}+i d_{3z^2-r^2}$ pairing, the gapless surface Majorana modes of the former get localized at the eight corners, thus yielding an \emph{intrinsic} third-order topological superconductor (TOTSC). A cousin $d_{xy}+id_{3z^2-r^2}$ pairing also accommodating eight corner Majorana modes, by virtue of breaking the cubic symmetry, in contrast, yields an \emph{extrinsic} TOTSC. We identify a doped octupolar (topological or trivial) Dirac insulator as a suitable platform to sustain such unconventional superconductors, realized from an intraunit cell pairing. Finally, we argue that the proposed TOTSC can be experimentally realizable in NaCl and other structurally similar compounds under high pressure.
\end{abstract}

\maketitle

\emph{Introduction}.~Localized Majorana zero modes are of paramount importance for braiding and non-Abelian statistics, and their applications in topological quantum computation~\cite{nayaketal:review, beenakker:review, oregoppen:review}. For these purposes, one-dimensional quantum nanowires offer a great potential as they can host topologically robust endpoint Majorana zero modes at low temperatures: a hallmark of the traditional bulk-boundary correspondence. Nonetheless, its recently discovered higher-order generalization manifesting through robust gapless modes localized on even lower-dimensional boundaries, such as corners and hinges~\cite{benalcazar2017, benalcazar-prb2017, song2017, schindler2018, trifunovic2017, gong2018, calugaru2019, fulga2019, szabo2020, surgoswami2021, bansil2021, gangchen2021, trauzettel2021}, when extended to the territory of neutral Bogoliubov-de Gennes (BdG) quasiparticles, boosts in this regard the prominence of higher-dimensional higher-order topological superconductors (HOTSCs)~\cite{wang-lin-hughes-HOTSC, wu-yan-huang-HOTSC, wang-liu-lu-zhang-HOTSC, liu-he-nori-HOTSC,  Klinovaja-HOTSC, yan-HOTSC, zhu-HOTSC, pan-yang-chen-xu-liu-liu-HOTSC, ghorashi-HOTSC, fulga-HOTSC-1, broyrantiunitary, trauzettel-HOTSC, bjyang-HOTSC-1, dassarma-HOTSC, srao-HOTSC, bomantara-HOTSC, broysoloHOTSC2020, kheirkhah2020, sigrist2020, thomale2020PRX, tiwari2020, ghoshnagsaha2021, shen2021, luopanliu2021}. For example, in contrast to conventional (or first-order) topological $p+ip$ and $d+id$ pairings, supporting one-dimensional Majorana edge modes, a two-dimensional $p+id$ HOTSC hosts four pointlike corner localized Mojorana modes~\cite{wang-lin-hughes-HOTSC, broysoloHOTSC2020}. However, thus far the proposed three-dimensional (3D) HOTSCs only encompass Majorana hinge modes, while the mechanism and the platforms for the realizations of 3D corner Majorana modes remained elusive. In this Letter, we therefore venture the following set of questions, and provide definite answers to them. (1) What is the underlying pairing symmetry of 3D HOTSCs that supports corner Majorana modes? (2) What are the suitable material platforms where such pairings can be realized?

\begin{figure}[t!]
\subfigure[]{\includegraphics[width=0.45\linewidth]{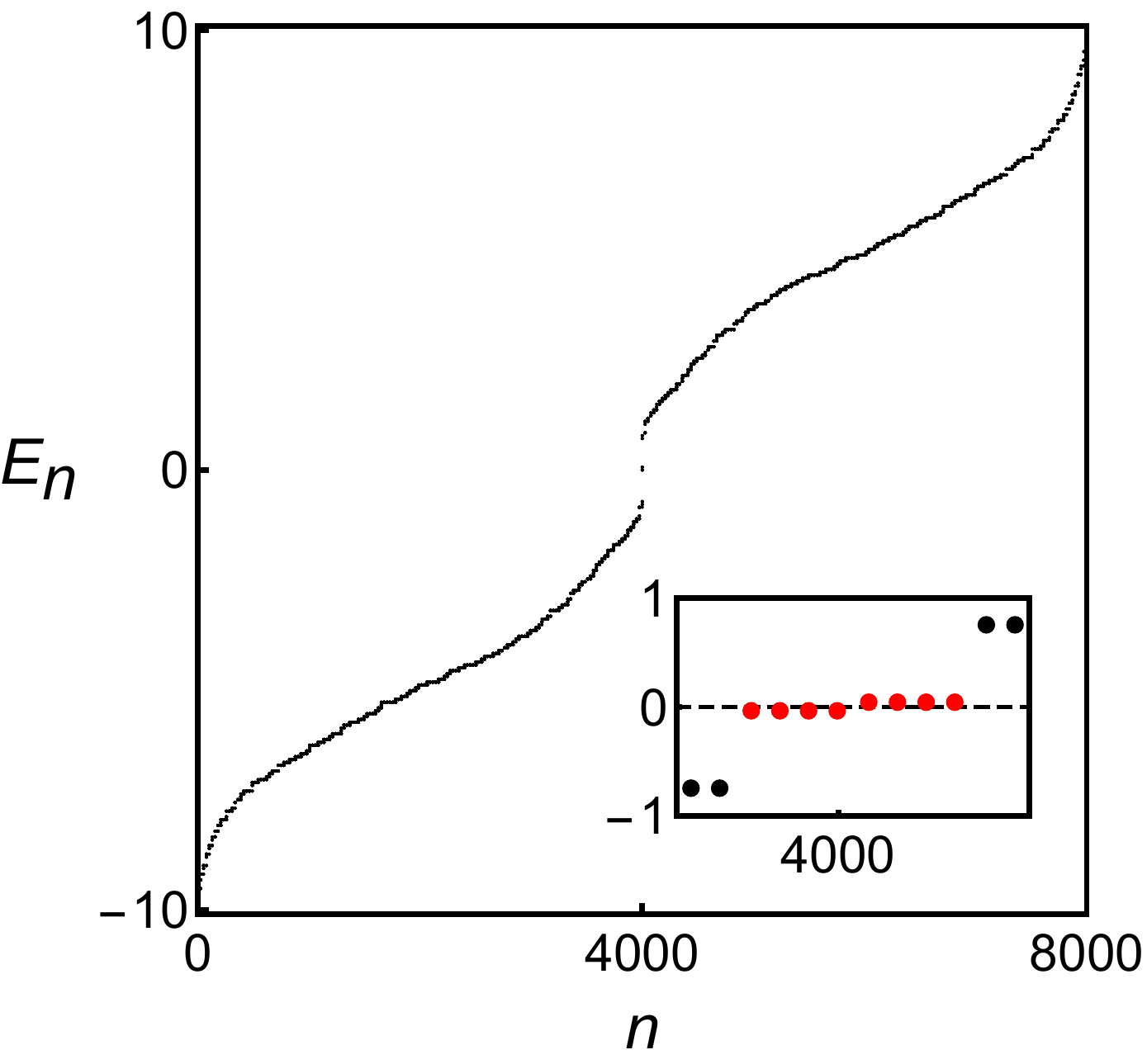}}
\subfigure[]{\includegraphics[width=0.45\linewidth]{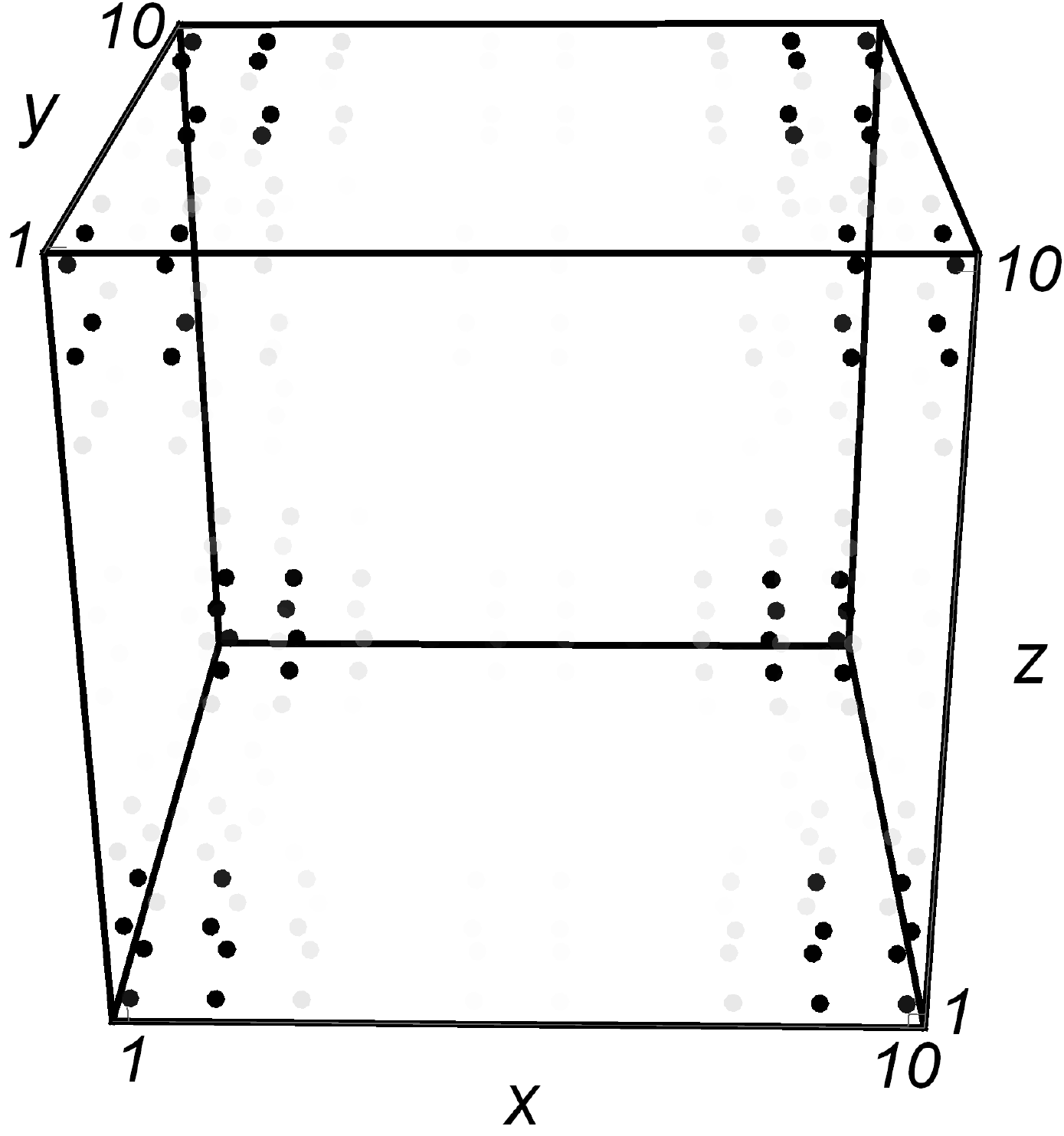}}%
\includegraphics[width=0.075\linewidth]{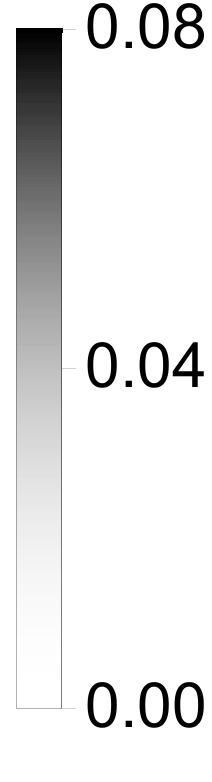}
\caption{(a) Eigenvalue spectra ($E_n$) for an intrinsic TOTSC, realized around a Fermi surface, on a cubic lattice. Inset: Eight near (due to finite system size) zero-energy corner modes (red dots), well separated from nearby bulk states (black dots). (b) Local density of states for the zero energy states in (a), displaying sharp localization around the corners in the $\langle 111 \rangle$ directions. These results remain qualitatively unchanged in the presence of a small $s$-wave component, and for the local pairing shown in Eq.~(\ref{eq:localoctu}) in an octupolar (topological or trivial) Dirac insulator (doped or undoped)~\cite{supplementary}. The linear dimension of the system is $L=10$ in each direction, and $t_1=t_0=\Delta_1=\Delta_2=m_0/2=1$ and $\Delta_s=0$ in Eq.~(\ref{eq:OctupolLattice}).
}~\label{Fig:IntrinsicTOTSC}
\end{figure}

\begin{figure}[t!]
\subfigure[]{\includegraphics[width=0.45\linewidth]{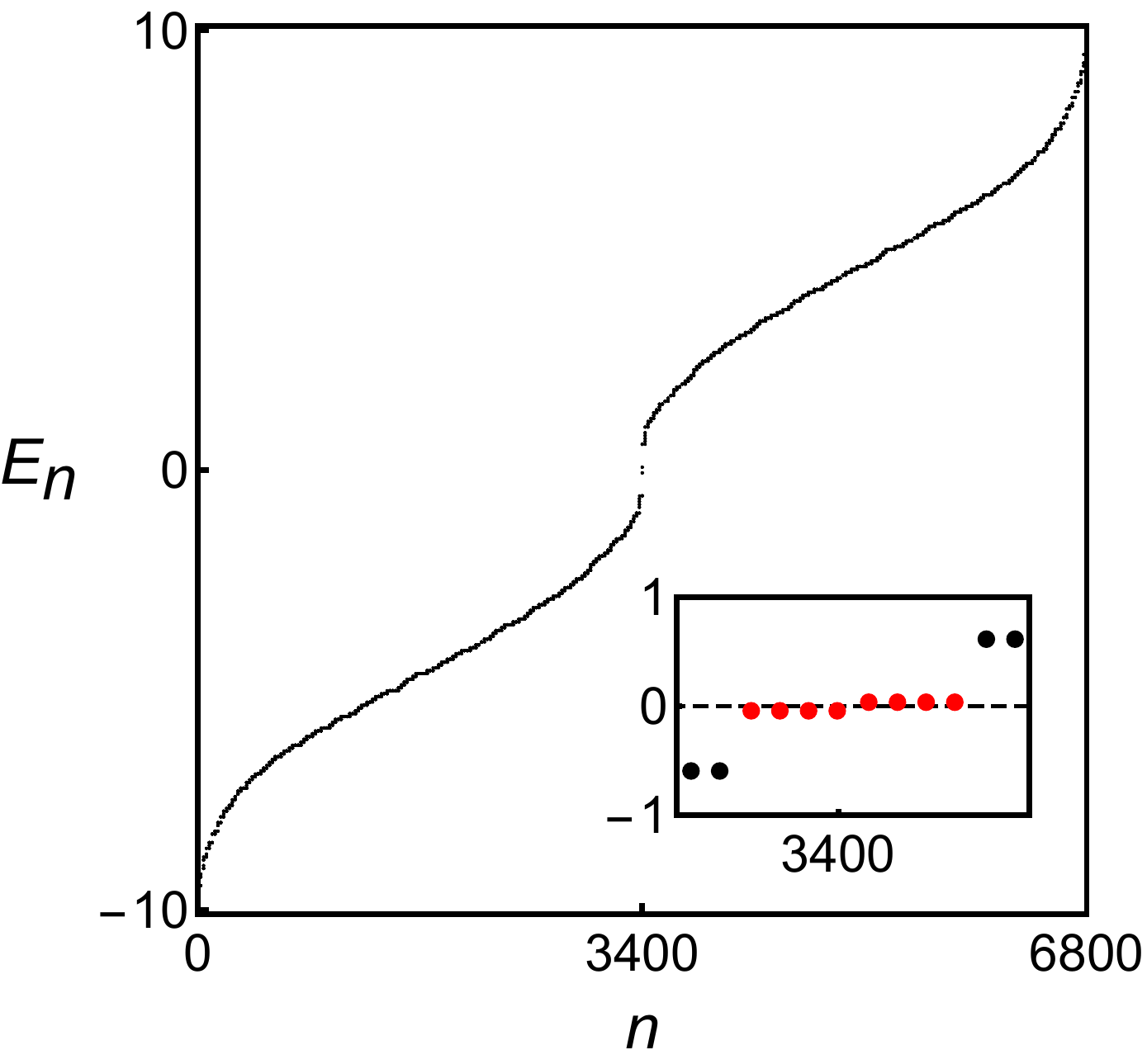}}
\subfigure[]{\includegraphics[width=0.45\linewidth]{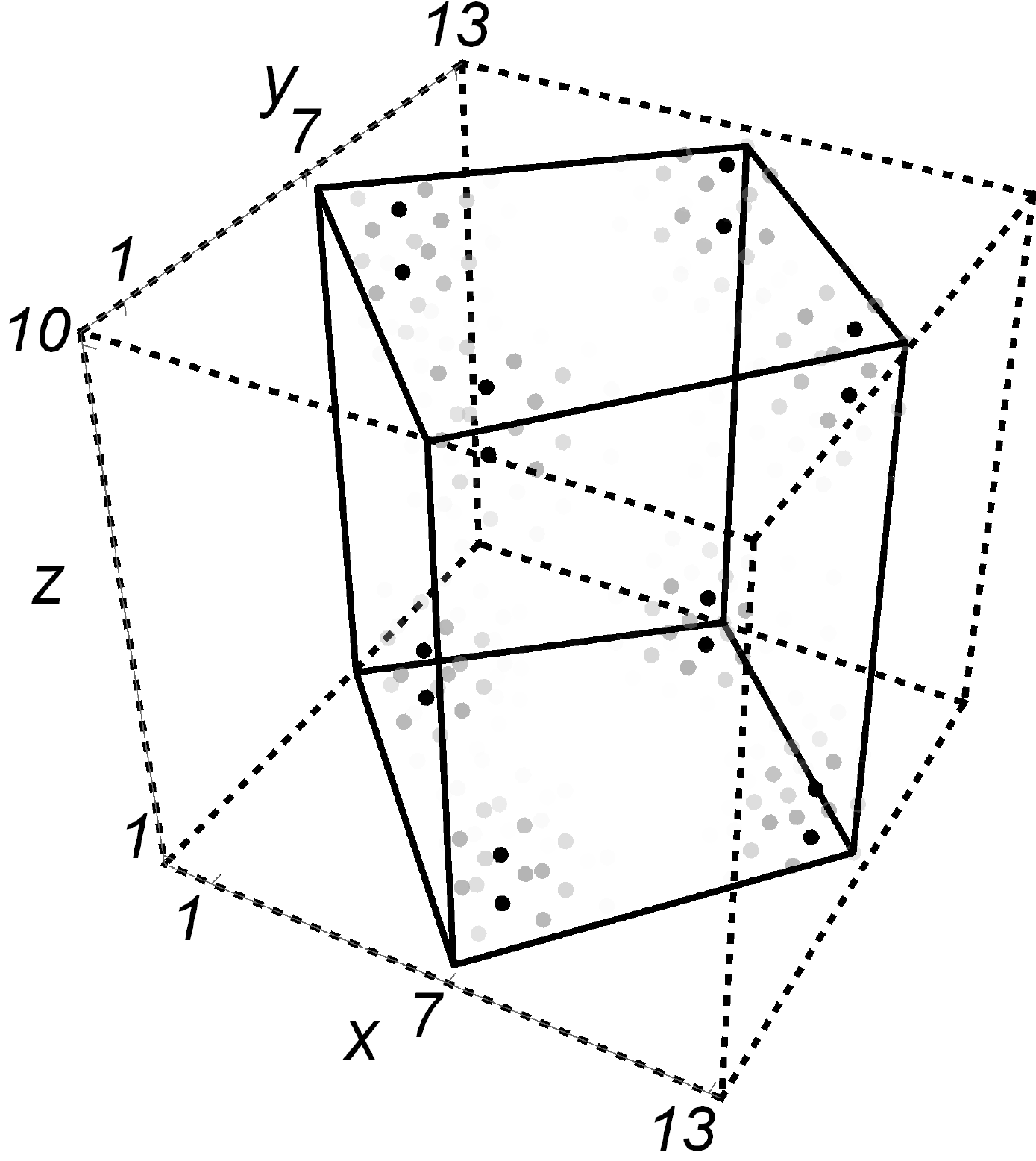}}%
\includegraphics[width=0.075\linewidth]{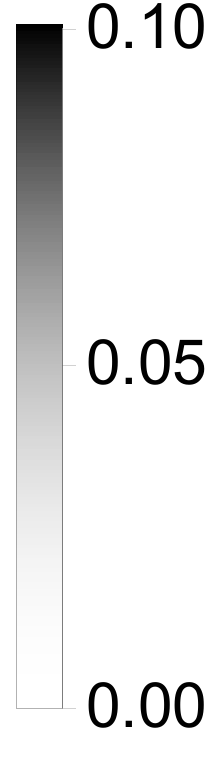}
\caption{(a) Eigenvalue spectra ($E_n$) for an extrinsic TOTSC, realized around a Fermi surface, on a cubic lattice (dashed cube), cleaved in such a way (solid cube) that eight corners are now placed at the four side centers on each of the two $xy$ planes. Inset: Eight near zero-energy corner modes (red dots), well separated from nearby bulk states (black dots). (b) Local density of states for the zero-energy states, displaying sharp corner localization. These results remain qualitatively unchanged in the presence of a small $s$-wave component, and for the local pairing shown in Eq.~(\ref{eq:localoctu}) in an octupolar Dirac material~\cite{supplementary}. The linear dimensions of the system are $L_x=13$, $L_y=13$, and $L_z=10$ in the $x$, $y$ and $z$ directions, respectively. The parameter values are the same as in Fig.~\ref{Fig:IntrinsicTOTSC}.
}~\label{Fig:ExtrinsicTOTSC}
\end{figure}

\begin{figure}[t!]
\includegraphics[width=0.485\linewidth]{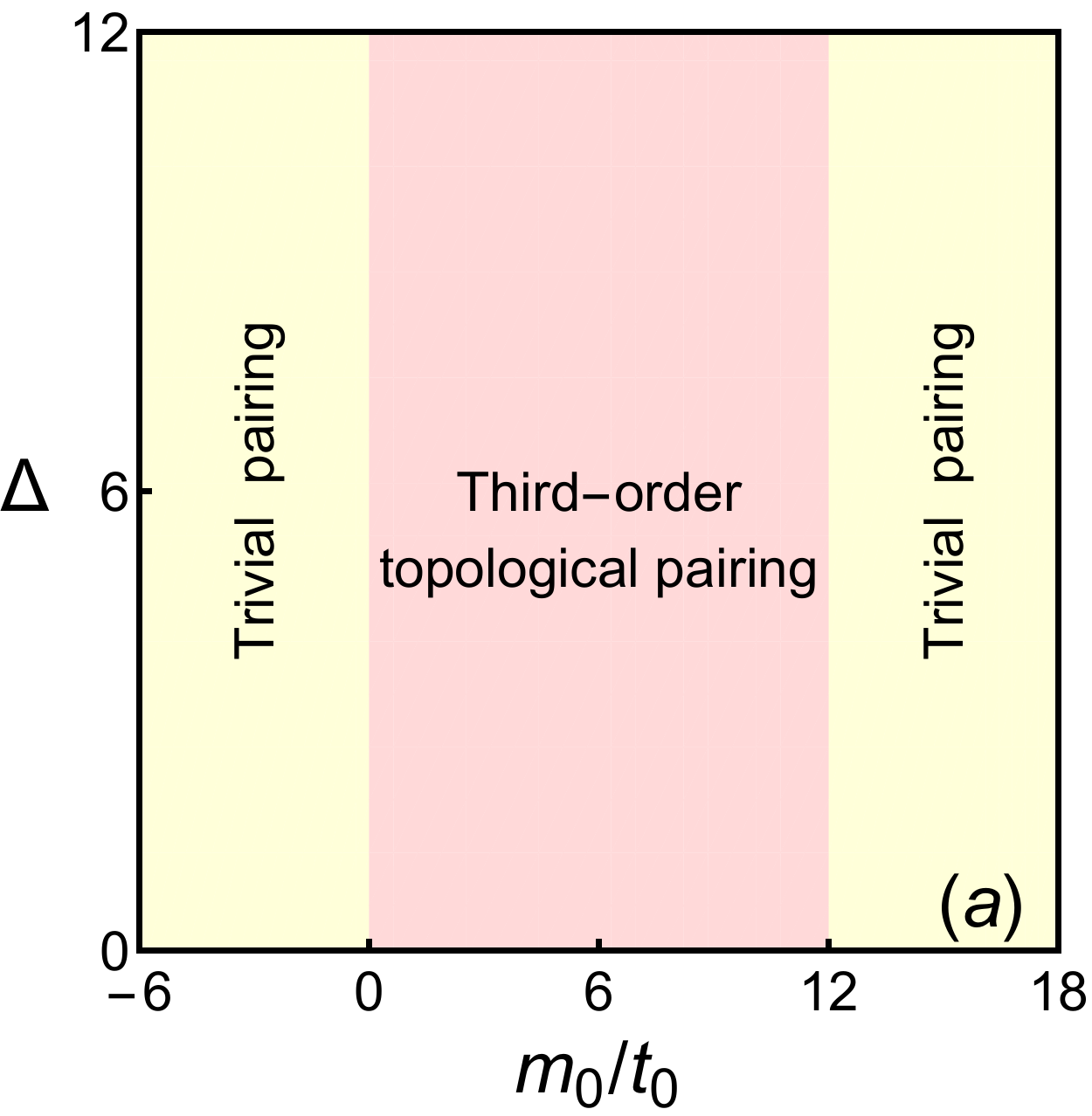}
\includegraphics[width=0.485\linewidth]{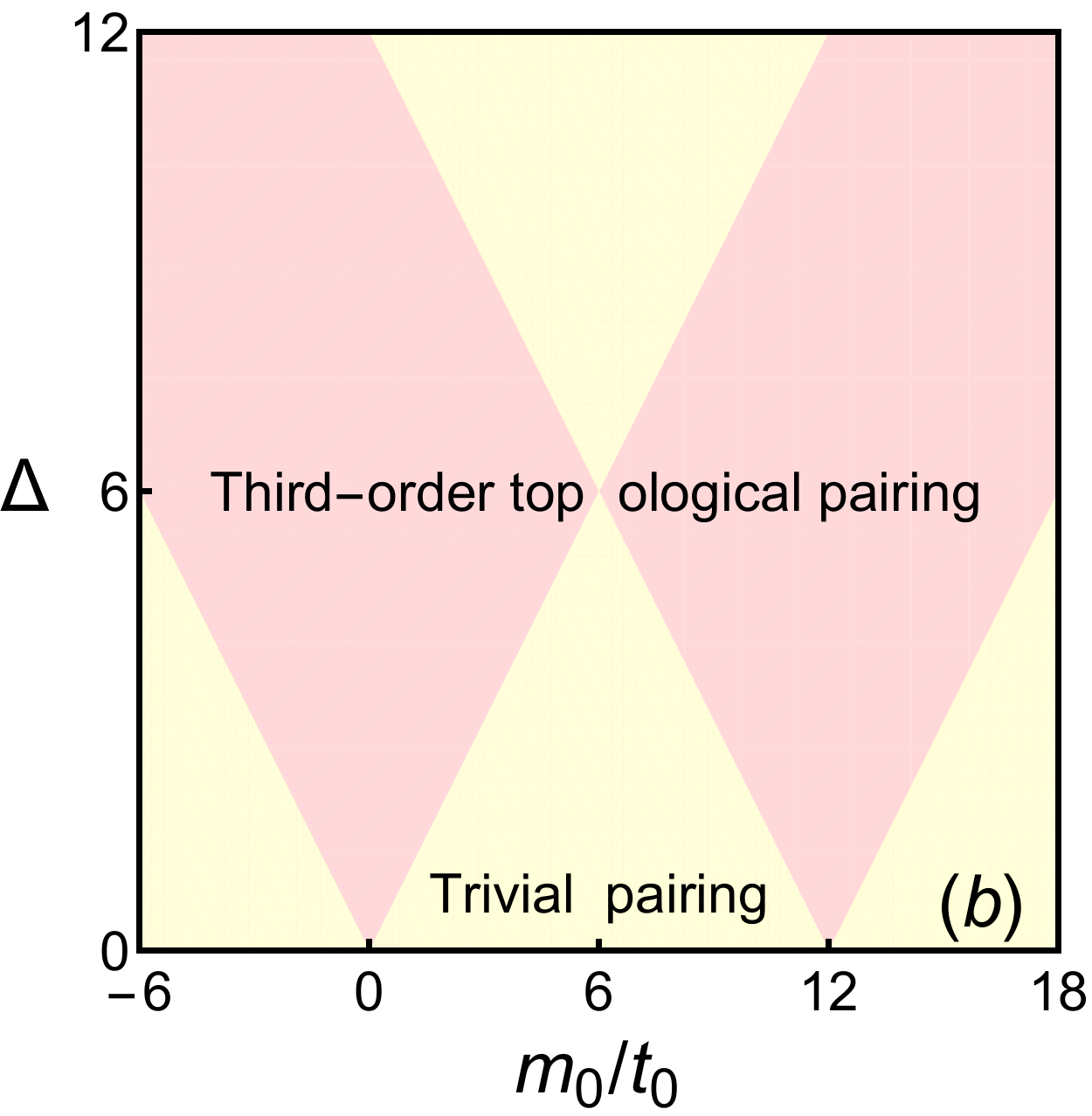}
\caption{Phase diagrams of TOTSCs (always supporting eight corner Majorana modes) for (a) lattice regularized BdG Hamiltonian and (b) local pairing in Eq.~(\ref{eq:localoctu}) in an octupolar Dirac insulator for $t_1=1$. For intrinsic (extrinsic) TOTSC $Q_{xyz}=0.5 \; (0.0)$. Trivial pairing does not support any corner modes and $Q_{xyz}=0$ therein. In (a) $\Delta_1=\Delta_2=\Delta$, while in (b) $\Delta_1=\Delta_2=1.0$ and $\Delta$ denotes amplitude of the local pairing in Eq.~(\ref{eq:localoctu}). The octupolar Dirac insulator is topological (trivial) for $0<m_0/t_0<12$ ($m_0/t_0<0$ and $m_0/t_0>12$).
}~\label{Fig:Phasediagrams}
\end{figure}

\emph{Key results}.~Here, we identify two candidate mixed-parity time-reversal symmetry breaking octupolar pairings, each of which supports eight zero-energy Majorana corner modes in suitably cleaved cubic crystals [Figs.~\ref{Fig:IntrinsicTOTSC} and~\ref{Fig:ExtrinsicTOTSC}]. Specifically, we show that when an odd-parity spin-triplet isotropic $p$-wave pairing (analog of the B-phase of $^3$He) coexists with an even-parity singlet $d_{x^2-y^2}+id_{3z^2-r^2}$ pairing, the resulting mixed parity superconducting state supports eight Majorana corner modes. This pairing is a prototypical example of octupolar pairing in a cubic system, transforming under the irreducible $E_g$ representation. It breaks the time-reversal symmetry, but preserves the cubic symmetry. Thus $p \oplus (d_{x^2-y^2}+id_{3z^2-r^2})$ pairing stands as an \emph{intrinsic} HOTSC~\cite{comment:symbol}. A cousin $p \oplus (d_{xy}+id_{3z^2-r^2})$ pairing, transforming under the mixed $T_{2g}$ and $E_g$ representations, although supporting eight corner Majorana modes, breaks the cubic symmetry. It thus stands as an \emph{extrinsic} HOTSC. Since pointlike corner Majorana modes with dimensionality $d_B=0$ in three dimensions ($d=3$) are characterized by the codimension $d_c=d-d_B=3$, these two paired states represent third-order topological superconductors (TOTSCs). They can be realized around an underlying Fermi surface  with an additional two-fold sublattice degeneracy besides the conventional Kramers degeneracy. The corner Majorana modes are stable even in the presence of a \emph{weak} $s$-wave pairing that gets induced naturally in the presence of dominant $d$-wave pairings. We identify a doped octupolar Dirac insulator (defined later) as a suitable platform where such unconventional pairings stem from a \emph{unique} fully gapped local pairing. While an intrinsic TOTSC possesses a quantized octupolar moment $Q_{xyz}=0.5$, for an extrinsic TOTSC $Q_{xyz}=0$. See the phase diagrams in Fig.~\ref{Fig:Phasediagrams}. Finally, the proposed TOTSC may be experimentally realizable in NaCl and structurally similar compounds InTe, SnAs and SnSb under high pressure~\cite{Stepanov-1979, SC:NaClStructure1, SC:NaClStructure2, SC:NaClStructure3}.

\emph{HOTSCs around Fermi surface}.~The effective single particle BdG Hamiltonian in the presence of $p \oplus (d_{\alpha}+id_{3z^2-r^2})$ pairings, with $\alpha=x^2-y^2$ and $xy$, around the Fermi surface (FS), possessing Kramers and two-fold sublattice degeneracy reads
\allowdisplaybreaks[4]
\begin{eqnarray}~\label{eq:octupoleFS}
H^{\rm FS}_{\rm octu} &=& \left( \frac{\vec{k}^2}{2 m_\ast} -\mu \right) \Gamma_{300}
                       + \Delta_p \; \sum^3_{j=1} \; \frac{k_{_j}}{k_{_F}} \; \Gamma_{13j} \nonumber \\
										  &+& \Delta_1 \; d_1 (\vec{k}) \; \Gamma_{110} + \Delta_2 \; d_2 (\vec{k}) \; \Gamma_{200} + \Delta_s \Gamma_{100},
\end{eqnarray}
where $\Gamma_{\mu \nu \rho}=\eta_\mu \tau_\nu \sigma_\rho$. Three sets of Pauli matrices $\{ \eta \}$, $\{ \tau \}$, and $\{ \sigma \}$ respectively act on the Nambu or particle-hole, sublattice, and spin or Kramers indices, $m_\ast$ is the effective mass, $\mu$ is the chemical doping, and $k_{_F}$ is the Fermi momentum. Throughout we consider $m_\ast, \mu>0$, such that the pairing of sharp normal state quasiparticles takes place around a Fermi surface. We are then in the weak-coupling regime. The triplet $p$-wave pairing with amplitude $\Delta_p$ is odd under parity $\vec{k} \to -\vec{k}$, while it preserves the time-reversal symmetry. The two components of the cubic $d$-wave pairings (with explicit forms defined below) are even under parity, i.e., $d_{1,2}(-\vec{k})=d_{1,2}(\vec{k})$. But the component $d_2(\vec{k})$ is odd under the reversal of time, generated by ${\mathcal T}=\Gamma_{002} {\mathcal K}$, where ${\mathcal K}$ is the complex conjugation and ${\mathcal T}^2=-1$. In addition, we also include an $s$-wave pairing with amplitude $\Delta_s$, which preserves the time reversal symmetry and gets naturally induced in the presence of a $d$-wave pairing, as both pairing channels are even under parity. The above effective single particle Hamiltonian enjoys the particle-hole symmetry, generated by the antiunitary operator $\Theta=\Gamma_{202} {\mathcal K}$ with $\Theta^2=+1$ and $\Theta H^{\rm FS}_{\rm octu} \Theta^{-1}=-H^{\rm FS}_{\rm octu}$.

In the absence of $d$- and $s$-wave pairings, $H^{\rm FS}_{\rm octu}$ describes a fully gapped isotropic odd-parity $p$-wave pairing (class DIII). As such, it supports two copies of gapless Majorana states on all six surfaces of a cubic crystal, irrespective of its specific cut~\cite{supplementary}. When only the $d$-wave pairings are included, all the matrices appearing in $H^{\rm FS}_{\rm octu}$ mutually anticommute. Since then $H^{\rm FS}_{\rm octu}$ involves six mutually anticommuting matrices, their minimal dimensionality has to be eight, which in turn demands an additional two-fold sublattice degeneracy of the Fermi surface. We now address the role of the $d$-wave pairings for the realization of Majorana corner modes.

\emph{Intrinsic TOTSC}.~From five possible cubic $d$-wave pairings, one can construct only one combination with
\begin{equation}\label{eq:intTOTSC}
d_1(\vec{k})=\frac{\sqrt{3}}{2 k^2_{_F}} (k^2_x-k^2_y), \:\:\:
d_2(\vec{k})=\frac{1}{2 k^2_{_F}} (2k^2_z-k^2_x-k^2_y)
\end{equation}
that preserves the cubic symmetry, but breaks the time-reversal symmetry. The resulting $d_{x^2-y^2}+id_{3z^2-r^2}$ state is an octupolar pairing and supports eight Majorana-Weyl nodes at $\pm k_x=\pm k_y =\pm k_z=k_{_F}/\sqrt{3}$ (in the absence of other superconducting orders). Even though both $d$-wave components transform under the irreducible doublet $E_g$ representation of the cubic or $O_h$ point group, their amplitudes in Eq.~(\ref{eq:octupoleFS}) are set to be different, since these two pairings cannot be transformed into each other by an arbitrary SO(3) rotation. Nonetheless, their transition temperatures are the same, as expected~\cite{sigrist-ueda-rmp, roy-ghorashi-foster-nevidomskyy}.

In the presence of such octupolar pairing, the gapless surface states of the isotropic $p$-wave pairing get partially gapped, since all the involved matrices in Eq.~(\ref{eq:octupoleFS}) then mutually anticommute. In other words, the $d_{x^2-y^2}+id_{3z^2-r^2}$ pairing acts as a mass for gapless surface Majorana fermions of the $p$-wave superconductor. However, such a BdG Wilson-Dirac mass vanishes along the high-symmetry eight body-diagonal $\langle 111 \rangle$ directions. As a result, the surface states of isotropic $p$-wave pairing are left gapless only at eight corners of a cubic crystal cleaved so that they are placed at $(\pm 1, \pm 1, \pm 1)L/2$, where $L$ is the linear dimension of the system in each direction, see Fig.~\ref{Fig:IntrinsicTOTSC}. The resulting $p \oplus (d_{x^2-y^2}+id_{3z^2-r^2})$ pairing therefore stands as an \emph{intrinsic} TOTSC that supports eight zero-energy Majorana corner modes. On the other hand, when $\Delta_2=0$, the $xy$ surfaces and four hinges along the $z$ direction host gapless Majorana modes, and we realize a second-order topological superconductor~\cite{supplementary}.

\emph{Extrinsic TOTSC}.~Another octupolar pairing with
\begin{equation}\label{eq:extTOTSC}
d_1(\vec{k})=\frac{\sqrt{3}}{k^2_{_F}} (k_x k_y), \:\:\:
d_2(\vec{k})=\frac{1}{2 k^2_{_F}} (2k^2_z-k^2_x-k^2_y)
\end{equation}
that also supports eight Majorana Weyl nodes at $(\pm \sqrt{2},0,\pm 1)k_{_F}/\sqrt{3}$ and $(0,\pm \sqrt{2},\pm 1)k_{_F}/\sqrt{3}$ (in the absence of any other pairings), partially gaps out the surface Majorana modes of the isotropic $p$-wave superconductor. Such an octupolar pairing leaves eight corners gapless, which, as dictated by the $d_{xy}$ pairing component in Eq.~(\ref{eq:extTOTSC}), are pinned at the four side centers on each of the two $xy$ planes in real space, see Fig.~\ref{Fig:ExtrinsicTOTSC}. The above two components of the $d$-wave pairings respectively transform under the $T_{2g}$ and $E_g$ representations, thereby breaking the cubic symmetry and the corresponding  two amplitudes in Eq.~(\ref{eq:octupoleFS}) are generically different. The resulting mixed-parity $p \oplus (d_{xy}+i d_{3z^2-r^2})$ pairing then stands as an \emph{extrinsic} TOTSC. Once again if we switch off the $d_{3z^2-r^2}$-wave pairing, a second-order topological superconductor is realized~\cite{supplementary}.

\emph{Induced $s$-wave pairing}. Now we address the impact of the induced $s$-wave component on the fully gapped TOTSC. For a small amplitude of such parasitic $s$-wave pairing the spectra of BdG quasiparticles remain fully gapped, and the system continues to support eight localized corner Majorana modes~\cite{supplementary}. However, beyond a critical amplitude of the $s$-wave pairing, which for the intrinsic (int) and extrinsic (ext) TOTSCs are respectively
\begin{equation}
\Delta^{\ast, {\rm int}}_{s}=\Delta_p, \:\:\:
\Delta^{\ast, {\rm ext}}_{s}=\sqrt{\Delta^2_p+\Delta^2_1/3}\;,
\end{equation}
the fully gapped paired state becomes topologically trivial and thus no longer supports corner modes.

\emph{Lattice model and numerical results}. To anchor the above outlined key results, next we perform a numerical analysis on a cubic lattice. The lattice-regularized Hamiltonian  corresponding to Eq.~(\ref{eq:octupoleFS}), which, pending the representation of the $\Gamma$ matrices, also describes the octupolar Dirac insulator (defined below) and the realization of the TOTSC therein, reads~\cite{nag-juricic-roy:2021}
\allowdisplaybreaks[4]
\begin{align}~\label{eq:OctupolLattice}
&H^{\rm lat}_{\rm octu} = t_1 \sum_{j=1,2,3} \sin(k_j a) \Gamma_j + \Gamma_4\; m_1^{\rm lat}({\bf k})\nonumber\\
&- \Delta_1 \; \Gamma_5 \sqrt{3} \; d^{\rm lat}_1(\vec{k})- \Delta_2 \; \Gamma_6\; d^{\rm lat}_2(\vec{k}) + \Delta_s \tilde{\Gamma}.
\end{align}
Here $a$ is the lattice spacing and $m_1^{\rm lat}({\bf k})= m_0-6t_0+ 2t_0 [\cos(k_x a)+\cos(k_y a)+\cos(k_z a)]$ is the first-order Dirac mass. For intrinsic and extrinsic TOTSCs $d^{\rm lat}_1(\vec{k})=\cos(k_x a) -\cos (k_y a)$ and $\sin(k_x a)\sin(k_y a)$, respectively, while $d^{\rm lat}_2(\vec{k})=2 \cos(k_z a) -\cos(k_x a) -\cos (k_y a)$ in both cases. Here $j=1,2$ and $3$ correspond to $x$, $y$ and $z$, respectively. Comparing with Eq.~(\ref{eq:octupoleFS}), we find the following correspondences among the matrices $\Gamma_{j}=\Gamma_{13j}$ for $j=1,2,3$, $\Gamma_4=\Gamma_{300}$, $\Gamma_5=\Gamma_{110}$, $\Gamma_6=\Gamma_{200}$ and $\tilde{\Gamma}=\Gamma_{100}$. When expanded around the $\Gamma=(0,0,0)$ point of the cubic Brillouin zone, for example, $H^{\rm lat}_{\rm octu}$ takes the form of $H^{\rm FS}_{\rm octu}$ with $m_\ast=(2 t_0 a^2)^{-1}$, $\mu=m_0$, $k_{_F}=a^{-1}$, and $\Delta_p=t_1$. We implement the above tight binding model on a cubic lattice with open boundary condition and numerically diagonalize it for different cuts of the crystal. The results are displayed in Figs.~\ref{Fig:IntrinsicTOTSC} and~\ref{Fig:ExtrinsicTOTSC}. Eight zero energy Majorana corner modes are found when $0<m_0/t_0<12$. On the other hand, for $m_0/t_0<0$ and $m_0/t_0>12$, the paired state is topologically trivial~\cite{supplementary}. In the following, we identify the octupolar Dirac insulator as a suitable platform for the realization of the TOTSC and the corresponding Majorana corner modes.

\emph{Octupolar Dirac insulators}.~The lattice model for an octupolar Dirac insulator takes the form shown in Eq.~(\ref{eq:OctupolLattice}) when $\Delta_s=0$, with eight-component mutually anticommuting Hermitian $\Gamma$ matrices now given by $\Gamma_j = \beta_1 \tau_1 \sigma_j$ for $j=1,2,3$, $\Gamma_4 = \beta_1 \tau_3 \sigma_0$, $\Gamma_5=\beta_1 \tau_2 \sigma_0$, and $\Gamma_6=\beta_2 \tau_0 \sigma_0$. Three sets of Pauli matrices $\{ \sigma \}$, $\{ \tau \}$ and $\{ \beta \}$ respectively act on the spin ($\uparrow, \downarrow$), parity ($\pm$) and sublattice (A,B) indices. The Hamiltonian is invariant under a composite ${\mathcal P}{\mathcal T}$ symmetry, where ${\mathcal T}=(\beta_0 \tau_0 \sigma_2) {\mathcal K}$, ${\mathcal P}=\beta_1 \tau_3 \sigma_0$, and under ${\mathcal P}$: $\vec{k} \to - \vec{k}$. Here ${\mathcal T}$ and ${\mathcal P}$ respectively play the role of time-reversal and parity operators, with $({\mathcal T} {\mathcal P})^2=-1$. Furthermore, the Hamiltonian is invariant under an additional parity operator ${\mathcal P}^{\prime}=\beta_2 \tau_1 \sigma_0$ and ${\mathcal P}^\prime: \vec{k} \to -\vec{k}$, and enjoys a unitary particle-hole or chiral symmetry, generated by $\Gamma_7=\beta_3 \tau_0 \sigma_0$. Even though the above model for $0< m_0/t_0<12$ supports a topological octupolar insulator with charged corner modes, here we consider the trivial regimes, $m_0/t_0<0$ and $m_0/t_0>12$. The normal state then does not support any topological boundary modes. Therefore, the appearance of Majorana bound states can solely be attributed to pairing, which we discuss next.

To select the pairing realizing the TOTSC in an octupolar insulator, we first notice that the system supports 28 (the number of purely imaginary eight-component Hermitian matrices) local (onsite or intra-unit) cell pairings, due to the Pauli exclusion principle. To capture all the pairings in a unified framework we Nambu-double the original eight-component spinor, and absorb the unitary part of the time-reversal operator (${\mathcal T}$) in the hole part of the Nambu spinor. In such a basis the octupolar Dirac insulator takes the form shown in Eq.~(\ref{eq:OctupolLattice}), with sixteen-dimensional $\Gamma$ matrices taking the explicit forms
\allowdisplaybreaks[4]
\begin{eqnarray}~\label{eq:ODI-gamma-matrices}
\Gamma_1 &=& \eta_3 \beta_1 \tau_1 \sigma_1,\:\: \Gamma_2 = \eta_3 \beta_1 \tau_1 \sigma_2,\:\: \Gamma_3 = \eta_3 \beta_1 \tau_1 \sigma_3,
\nonumber \\
\Gamma_4 &=& \eta_3 \beta_1 \tau_3 \sigma_0, \:\: \Gamma_5 = \eta_0 \beta_1 \tau_2 \sigma_0, \:\: \Gamma_6=\eta_0 \beta_2 \tau_0 \sigma_0.
\end{eqnarray}
The chemical potential term is given by $-\mu (\eta_3 \beta_0 \tau_0 \sigma_0)$.

A  local pairing (with a constant amplitude) supporting  Majorana corner modes satisfies the following algebraic constraints. It \emph{anticommutes} with the Dirac kinetic energy (proportional to $t_1$) and \emph{commutes} with the first-order Dirac mass~\cite{pairingmassexplanation}. The paired state then represents a fully gapped topological pairing with two-dimensional dispersive massless Majorana modes occupying all six surfaces of a cubic crystal, when $\Delta_1=\Delta_2=0$. In addition, the paired state must also simultaneously \emph{anticommute} with two higher-order Wilson-Dirac \emph{insulating} masses (proportional to $\Gamma_5$ and $\Gamma_6$), such that surface states get partially gapped, leaving eight corners gapless. Only one pairing satisfies all these constraints~\cite{supplementary}, for which the effective single-particle Hamiltonian is
\begin{equation}~\label{eq:localoctu}
H_{\rm octu}= \Delta \left( \eta_1 \cos \phi + \eta_2 \sin \phi \right) \beta_1 \tau_1 \sigma_0,
\end{equation}
where $\phi$ is the U(1) superconducting phase and $\Delta$ is the pairing amplitude. This pairing is a spin-singlet, but mixes even and odd parity bands, and two sublattices. We numerically diagonalize $H^{\rm lat}_{\rm octu}$ corresponding to the octupolar insulator in the presence of this pairing and find the eight zero-energy corner Majorana modes in a cubic system, cleaved according to the chosen form of $d^{\rm lat}_1(\vec{k})$, similar to Figs.~\ref{Fig:IntrinsicTOTSC} and~\ref{Fig:ExtrinsicTOTSC}, thus yielding a TOTSC. If, on the other hand, we set $\Delta_2=0$, the same paired state corresponds to a second-order topological superconductor with gapless hinge modes along the $z$ direction and surface states occupying the $xy$ surfaces~\cite{supplementary}.

These observations can be supported by projecting the above local pairing onto the Fermi surface using the band basis of the single-particle Hamiltonian in Eq.~\eqref{eq:OctupolLattice}, and neglecting the interband pairing components. The reduced Hamiltonian (after a suitable global unitary rotation) assumes the form of $H^{\rm FS}_{\rm octu}$ in Eq.~(\ref{eq:octupoleFS}), when expanded around the $\Gamma$ or $R$ point of the Brillouin zone. Furthermore, with appropriate choices of the insulating mass form factor $d^{\rm lat}_1(\vec{k})$ the same local pairing from Eq.~(\ref{eq:localoctu}) yields either intrinsic or extrinsic TOTSC~\cite{supplementary}. Therefore, the local pairing $H_{\rm octu}$ \emph{imposes} a nontrivial octupolar topology when projected onto the Fermi surface, in spite of the parent insulating phase being trivial. These conclusions remain qualitatively unchanged when the normal state is a topological octupolar insulator.

\emph{Topological invariant}.~Intrinsic and extrinsic TOTSCs can be distinguished besides by symmetry, also in terms of a bulk topological invariant, the octupolar moment $Q_{xyz}$~\cite{hughes:octupolar, cho:octupolar, agarwala:octupolar}. To extract $Q_{xyz}$, we treat holelike excitations as independent particlelike excitations and compute
\begin{equation}
n={\rm Re} \left[ -\frac{i}{2 \pi} {\rm Tr} \left( \ln \left\{ U^\dagger  \exp \left[ 2 \pi i \sum_{\bf r} \hat{q}_{xyz} ({\bf r}) \right]  U \right\} \right) \right],
\end{equation}
where $\hat{q}_{xyz} ({\bf r})= x y z \hat{n}({\bf r})/L^3$, $\hat{n}({\bf r})$ is the number operator at ${\bf r}=(x,y,z)$ of a periodic cubic system of linear dimension $L$ in each direction, and $U$ is constructed by columnwise arranging the eigenvectors for the negative energy states. The octupolar moment is defined as $Q_{xyz}=n-n_{\rm al}$ (modulo 1), where $n_{\rm al}=(1/2) \; \sum_{\bf r} x y z /L^3$ represents $n$ in the atomic limit and at half filling. We compute $Q_{xyz}$ for the lattice regularized BdG Hamiltonian and the local pairing in an octupolar Dirac insulator [Eqs.~(\ref{eq:OctupolLattice}) and ~(\ref{eq:localoctu})], which depending on the form factor $d^{\rm lat}_1(\vec{k})$ yields intrinsic or extrinsic TOTSC. While the octupolar moment is quantized $Q_{xyz}=0.5$ in an intrinsic TOTSC, $Q_{xyz}=0$ in an extrinsic TOTSC. In terms of the corner modes and $Q_{xyz}$, we construct cuts of the phase diagram for intrinsic and extrinsic TOTSCs in Fig.~\ref{Fig:Phasediagrams}.

\emph{Summary and discussions}.~We show that time-reversal symmetry breaking mixed parity octupolar $p \oplus (d_\alpha + i d_{3z^2-r^2})$ pairing supports eight corner localized Majorana modes in properly cleaved cubic crystals [Figs.~\ref{Fig:IntrinsicTOTSC} and ~\ref{Fig:ExtrinsicTOTSC}]. There are two such orders, representing intrinsic (for $\alpha=x^2-y^2$) and extrinsic (for $\alpha=xy$) TOTSCs. The corner modes can be detected by scanning tunneling microscopy, for example. We furthermore identify a doped octupolar (topological or trivial) Dirac insulator as the suitable material platform where such superconducting order can arise from local or on-site Cooper pairs. Remarkably, among all possible local pairings in this system, the unique pairing supporting the Majorana corner modes is also energetically most favored over a wide range of $m_0/t_0$, covering both topological and trivial Dirac insulating phases in the normal state~\cite{supplementary}. In addition, the TOTSC and its associated corner modes remain stable in the presence of a weak induced $s$-wave pairing.

Presently, NaCl is the only known candidate material for octupolar topological Dirac insulator~\cite{watanabe:arxiv2020} and it may be a superconductor under pressure with transition temperature $T_c \sim$ 2-7K~\cite{Stepanov-1979}. Nonetheless, structurally analogous binary compounds such as InTe, SnAs and SnSb under high pressure also show superconductivity with $T_c \sim$ 1-3K~\cite{SC:NaClStructure1, SC:NaClStructure2, SC:NaClStructure3}. Given that our analysis suggests that the doped octupolar Dirac insulator does not need to be topological to accommodate TOTSC, which is at the same time energetically most favorable topological pairing in this system~\cite{supplementary}, we expect that te topological nature of superconductivity in these materials will be scrutinized more thoroughly in the future. Our proposal should also stimulate the search for new octupolar Dirac materials. Indeed, a recent study~\cite{mao2021} reported possible candidate materials for the realization of the octupolar Dirac insulator in Ti$_4$XTe$_3$, with X=Pb, Sn. When doped, these materials will constitute an ideal platform to harbor TOTSCs.

\emph{Acknowledgments}. B.R. was supported by the startup grant from Lehigh University and thanks Andr\' as L. Szab\' o for useful discussions. V.J. acknowledges support of the Swedish Research Council (VR 2019-04735).

\emph{Note added}.~After completing this work we became aware of a study where proximity-induced TOTSC in doped third-order topological insulator with preexisting charged corner modes has been discussed~\cite{Nag:HOTSC2021}.


\begin{thebibliography}{}
\bibitem{comment:symbol} The ``$\oplus$" symbol indicates that pairing matrices in the odd and even parity channels fully anticommnute. Two $d$-wave components also always mutually anticommute as they break the time-reversal symmetry.
\end{thebibliography}
\end{document}